\documentclass[11pt]{article}
\usepackage{amssymb}
\usepackage{amsmath}
\usepackage{a4wide}
\usepackage{graphicx}

\newtheorem{theorem}{Theorem}

\newenvironment{proof}[1][Proof]{\textbf{#1.} }{\ \rule{0.5em}{0.5em}}
\makeatletter
\def\@removefromreset#1#2{\let\@tempb\@elt
     \def\@tempa#1{@&#1}\expandafter\let\csname @*#1*\endcsname\@tempa
     \def\@elt##1{\expandafter\ifx\csname @*##1*\endcsname\@tempa\else
    \noexpand\@elt{##1}\fi}     \expandafter\edef\csname cl@#2\endcsname{\csname cl@#2\endcsname}     \let\@elt\@tempb
     \expandafter\let\csname @*#1*\endcsname\@undefined}

\@removefromreset{equation}{section}

\@removefromreset{theorem}{section}
\makeatother
\input{tcilatex}

\begin{document}

\title{Class of bipartite quantum states satisfying the original Bell
inequality }
\author{Elena R. Loubenets \\
Applied Mathematics Department,\\
Moscow State Institute of Electronics and Mathematics, \\
Trekhsvyatitelskii per. 3/12, Moscow 109028, Russia}
\maketitle

\begin{abstract}
In a general setting, we introduce a new bipartite state property sufficient
for the validity of the perfect correlation form of the original Bell
inequality for any three bounded quantum observables. A bipartite quantum
state with this property does not necessarily exhibit perfect correlations.
The class of bipartite states specified by this property includes both
separable and nonseparable states. We prove analytically that, for any
dimension $d\geq 3,$ every Werner state, separable or nonseparable, belongs
to this class.
\end{abstract}

\section{Introduction}

The validity of Bell-type inequalities in the quantum case has been
intensively discussed in the literature from the fundamental publications of
J. S. Bell [1] and J. F. Clauser, M. A. Horne, A. Shimony and R. A. Holt
[2]. At present, Bell-type inequalities are widely used in quantum
information processing. However, from the pioneering paper of R. Werner [3]
up to now a general structure of bipartite quantum states not violating
Bell-type inequalities has not been well formalized. The recent results of
M. Terhal, A. C. Doherty and D. Schwab [4] represent a significant progress
in this direction but concern only the validity of CHSH\footnote{%
Abbreviation of Clauser-Horne-Shimony-Holt.}-form inequalities. Moreover,
the proof in [4] of one of its main results on CHSH-form inequalities (see
[4], theorem 2), specified for the case of discrete outcomes, cannot be
explicitly extended to a general spectral case.

A general structure of bipartite quantum states satisfying the original Bell
inequality for any three bounded quantum observables\footnote{%
In classical probability, the product expectation values satisfy the perfect
correlation form of the original Bell inequality for any three bounded
classical observables and any classical state (see appendix of [5], for the
proof). Recall that the original derivation of this inequality in [1] is
true only for dichotomic classical observables with values $\pm \lambda .$}
has not been analyzed in the physical and mathematical literature.

The original derivation of the perfect correlation form of the Bell
inequality in [1] is essentially based on the assumption of perfect
correlations whenever one and the same quantum observable is measured on
both "sides". However, for a bipartite quantum state, separable or
nonseparable, the condition on perfect correlations cannot be fulfilled for
every quantum observable. On the other hand, as we proved in a general
setting in [5], there exist\footnote{%
See [5], section 3.B.1, Eq. (49).} separable quantum states that satisfy the
perfect correlation form of the original Bell inequality for any three
bounded quantum observables and do not necessarily exhibit perfect
correlations. From the mathematical point of view, the Bell correlation
assumptions in [1] represent only sufficient but not necessary conditions
for a bipartite quantum state to satisfy the original Bell inequality.
Therefore, there must exist more general sufficient conditions.

In the present paper, we analyze the validity of the CHSH inequality and the
original Bell inequality in a general bipartite quantum case. We introduce a
new bipartite state property sufficient for the validity of the perfect
correlation form of the original Bell inequality for any three bounded
quantum observables. This state property is purely geometrical and is
associated with the existence for a bipartite quantum state of a special
type dilation\footnote{%
This dilation differs from dilations introduced in [4].} to an extended
tensor product Hilbert space. Satisfying the perfect correlation form of the
original Bell inequality for any three bounded quantum observables, a
bipartite quantum state with this property does not necessarily exhibit
perfect correlations.

We prove that every Werner state [3] on $\mathbb{C}^{d}\otimes \mathbb{C}%
^{d} $, $\forall d\geq 3,$ separable or nonseparable, has this property and,
therefore, satisfies the original Bell inequality for any three quantum
observables on $\mathbb{C}^{d}.$ In the two-qubit case, the original Bell
inequality holds for any separable Werner state.

\section{Source-operators and DSO states}

For a quantum state $\rho $ on a separable complex Hilbert space $\mathcal{H}%
\otimes \mathcal{H}$, possibly infinite dimensional\textit{,} let us
introduce self-adjoint trace class operators $T_{\blacktriangleright }$ and $%
T_{\blacktriangleleft }$ on $\mathcal{H}\otimes \mathcal{H}\otimes \mathcal{H%
}$, defined by the relations 
\begin{equation}
\mathrm{tr}_{\mathcal{H}}^{(2)}[T_{\blacktriangleright }]=\mathrm{tr}_{%
\mathcal{H}}^{(3)}[T_{\blacktriangleright }]=\rho ,\text{ \ \ \ }\mathrm{tr}%
_{\mathcal{H}}^{(1)}[T_{\blacktriangleleft }]=\mathrm{tr}_{\mathcal{H}%
}^{(2)}[T_{\blacktriangleleft }]=\rho .  \label{1}
\end{equation}%
Here, $\mathrm{tr}_{\mathcal{H}}^{(k)}[\cdot ],$ $k=1,2,3,$ denotes the
partial trace over the elements of $\mathcal{H}$ standing in the $k$-th
place in $\mathcal{H}\otimes \mathcal{H}\otimes \mathcal{H}$ and the lower
indices of $T_{\blacktriangleright }$, $T_{\blacktriangleleft }$ indicate
the direction of extension. Note that $T_{\blacktriangleright }$, $%
T_{\blacktriangleleft }$ are not necessarily positive\footnote{%
They also do not necessarily have symmetries specified for dilations in [4].}%
.

For concreteness,\emph{\ }we call any of dilations defined by (\ref{1})%
\textit{\ a\ source-operator\ for a bipartite state} $\rho .$ For any
bipartite state $\rho ,$\ source-operators $T_{\blacktriangleright }$, $%
T_{\blacktriangleleft }$\ exist\footnote{%
See [6] (sec. 2.1, proposition 1), for the proof.}. From (\ref{1}) it
follows that, for any source-operator $T,$ its trace $\mathrm{tr}[T]=1.$
Since any positive source-operator is a density operator, we refer to it as
a \textit{density source-operator} or a \textit{DSO}, for short.

The notion of a source-operator allows us to derive the following general
upper bounds\footnote{%
See [6] (sec. 2.2, proposition 3), for the proof.} for quantum product
averages in an arbitrary bipartite state $\rho $: 
\begin{eqnarray}
\left\vert \text{ }\mathrm{tr}[\rho (W_{1}\otimes W_{2})]-\mathrm{tr}[\rho
(W_{1}\otimes \widetilde{W}_{2})]\text{ }\right\vert &\leq
&||T_{\blacktriangleright }||_{1}\{1-\mathrm{tr}[\sigma
_{T_{\blacktriangleright }}^{(1)}(W_{2}\otimes \widetilde{W}_{2})]\text{ }\},
\label{2} \\
\left\vert \text{ }\mathrm{tr}[\rho (W_{1}\otimes W_{2})-\mathrm{tr}[\rho (%
\widetilde{W}_{1}\otimes W_{2})]\text{ }\right\vert &\leq
&||T_{\blacktriangleleft }||_{1}\{1-\mathrm{tr}[\sigma
_{T_{\blacktriangleleft }}^{(3)}(W_{1}\otimes \widetilde{W}_{1})]\text{ }\}.
\label{3}
\end{eqnarray}%
Here: (i) $W_{1},$ $\widetilde{W}_{1},$ $W_{2},$ $\widetilde{W}_{2}$ are any
bounded quantum observables on $\mathcal{H}$ with operator norms $||\cdot
||\leq 1;$ (ii) $T_{\blacktriangleright }$, $T_{\blacktriangleleft }$ are
any source-operators for a state $\rho $; (iii) $\left\Vert T\right\Vert
_{1}:=\mathrm{tr}[\left\vert T\right\vert ]$ is the trace norm of a
source-operator $T$ and $\sigma _{T}^{(j)}:=\frac{1}{||T||_{1}}\mathrm{tr}_{%
\mathcal{H}}^{(j)}[\left\vert T\right\vert ],$ $j=1,3,$ is the density
operator on $\mathcal{H}\otimes \mathcal{H}$ induced by $T.$

If, for a bipartite state $\rho $, there exists a density source-operator
then we call this $\rho $ \textit{a density source-operator state}\footnote{%
Any bipartite state that has an $(s_{a},s_{b})$-symmetric extension
(according to the terminology used in [4]) represents a DSO state. From the
other side, the corresponding symmetrization of a density source-operator
results in an (1,2) or (2,1) symmetric extension.}\emph{\ }or a \textit{DSO
state,} for short.\emph{\ }

For a density source-operator $R$, its trace norm $\left\Vert R\right\Vert
_{1}=1.$ Therefore, for a DSO state $\rho ,$ the bounds (\ref{2}), (\ref{3}%
), specified with the corresponding density source-operators $%
R_{\blacktriangleright }$ or $R_{\blacktriangleleft }$, take the form 
\begin{eqnarray}
\left\vert \text{ }\mathrm{tr}[\rho (W_{1}\otimes W_{2})]-\mathrm{tr}[\rho
(W_{1}\otimes \widetilde{W}_{2})]\text{ }\right\vert &\leq &1-\mathrm{tr}%
[\sigma _{R_{\blacktriangleright }}^{(1)}(W_{2}\otimes \widetilde{W}_{2})]%
\text{ }\},  \label{4} \\
\left\vert \text{ }\mathrm{tr}[\rho (W_{1}\otimes W_{2})]-\mathrm{tr}[\rho (%
\widetilde{W}_{1}\otimes W_{2})]\text{ }\right\vert &\leq &1-\mathrm{tr}%
[\sigma _{R_{\blacktriangleleft }}^{(3)}(W_{1}\otimes \widetilde{W}_{1})]%
\text{ }\},  \label{5}
\end{eqnarray}%
where $\sigma _{R}^{(j)}=\mathrm{tr}_{\mathcal{H}}^{(j)}[R].$

We introduce the following general statement on DSO states.

\begin{theorem}
A DSO state $\rho $ on $\mathcal{H}\otimes \mathcal{H}$ satisfies the
original CHSH inequality [2]: 
\begin{equation}
\left| \text{ }\mathrm{tr}[\rho (W_{1}\otimes W_{2}+W_{1}\otimes \widetilde{W%
}_{2}+\widetilde{W}_{1}\otimes W_{2}-\widetilde{W}_{1}\otimes \widetilde{W}%
_{2})]\right| \leq 2,  \label{6}
\end{equation}
for any bounded quantum observables\footnote{%
In case of an infinite dimensional $\mathcal{H},$ observables may be of any
spectral type.} $W_{1},$ $\widetilde{W}_{1}$, $W_{2},$ $\widetilde{W}_{2}$
on $\mathcal{H}$ with operator norms $\left\| \cdot \right\| \leq 1.$
\end{theorem}

\begin{proof}
Let a DSO\ state $\rho $ have a density source-operator $R_{%
\blacktriangleright }.$ Then, combining in the left-hand side of the
inequality (\ref{6}) the first term with the second while the third term
with the fourth and applying further (\ref{4}), we have 
\begin{eqnarray}
&&\left\vert \text{ }\mathrm{tr}[\rho (W_{1}\otimes W_{2}+W_{1}\otimes 
\widetilde{W}_{2}+\widetilde{W}_{1}\otimes W_{2}-\widetilde{W}_{1}\otimes 
\widetilde{W}_{2})]\right\vert  \label{7} \\
&\leq &\left\vert \mathrm{tr}[\rho (W_{1}\otimes W_{2}+W_{1}\otimes 
\widetilde{W}_{2})]\right\vert +\left\vert \mathrm{tr}[\rho (\widetilde{W}%
_{1}\otimes W_{2}-\widetilde{W}_{1}\otimes \widetilde{W}_{2})]\right\vert
\leq 2.  \notag
\end{eqnarray}%
If a state $\rho $ has a density source-operator $R_{\blacktriangleleft }$,
then we prove (\ref{6}) quite similarly - by combining in the left-hand side
of (\ref{6}) the first term with the third while the second term with the
fourth and applying further (\ref{5}).
\end{proof}

Any separable state is a DSO state\footnote{%
Let $\sum_{i}\alpha _{i}\rho _{i}\otimes \widetilde{\rho }_{i},$ $\alpha
_{i}>0,$ $\sum_{i}\alpha _{i}=1,$ be a separable representation of a
separable state. Then, for this state, $\sum_{i}\alpha _{i}\rho _{i}\otimes 
\widetilde{\rho }_{i}\otimes \widetilde{\rho }_{i}$ and $\sum_{i}\alpha
_{i}\rho _{i}\otimes \rho _{i}\otimes \widetilde{\rho }_{i}$ represent
density source-operators.}. In section 3.1, we present examples of
nonseparable DSO states. Notice that a nonseparable DSO state does not
necessarily admit a local hidden variable model in the sense formulated in
[3].

Consider now a generalized joint quantum measurement, with real-valued
outcomes $\lambda _{1},\lambda _{2}\in \Lambda \subseteq \lbrack -1,1]$ of
any spectral type and performed on a bipartite state $\rho .$ Let, under
this measurement, the joint probability that outcomes $\lambda _{1}$ and $%
\lambda _{2}$ belong to subsets $B_{1}$, $B_{2}\subseteq \Lambda ,$
respectively, have the form\footnote{%
See, for example, in [5].}: $\mathrm{tr}[\rho (M_{1}^{(a)}(B_{1})\otimes
M_{2}^{(b)}(B_{2}))],$ where $M_{1}^{(a)}$ and $M_{2}^{(b)}$ are positive
operator-valued (\textit{POV}) measures of both parties involved and
parameters $a,$ $b$ specify measurement settings of these parties. In the
physical literature, this type of a joint measurement is usually associated
with Alice and Bob names. Under an Alice/Bob joint measurement, specified by
a pair\footnote{%
Here, the first argument in a pair refers to a marginal measurement (say of
Alice) with outcomes $\lambda _{1}$, while the second argument - to a Bob
marginal measurement, with outcomes $\lambda _{2}.$} $(a,b)$ of measurement
settings, the expectation value $\langle \lambda _{1}\lambda _{2}\rangle
_{\rho }^{(a,b)}$ of the product $\lambda _{1}\lambda _{2}$ of outcomes has
the form 
\begin{equation}
\langle \lambda _{1}\lambda _{2}\rangle _{\rho }^{(a,b)}:=\int_{\Lambda
\times \Lambda }\lambda _{1}\lambda _{2}\mathrm{tr}[\rho
(M_{1}^{(a)}(d\lambda _{1})\otimes M_{2}^{(b)}(d\lambda _{2}))]=\mathrm{tr}%
[\rho (W_{1}^{(a)}\otimes W_{2}^{(b)})],  \label{8}
\end{equation}%
where $W_{1}^{(a)}:=\int_{\Lambda }\lambda _{1}M_{1}^{(a)}(d\lambda _{1})$
and $W_{2}^{(b)}:=\int_{\Lambda }\lambda _{2}M_{2}^{(b)}(d\lambda _{2})$ are
bounded quantum observables, representing on $\mathcal{H}$ marginal
measurements of Alice and Bob, respectively.

Due to theorem 1 and the representation (\ref{8}), we immediately derive
that a DSO state $\rho $ satisfies the CHSH inequality: 
\begin{equation}
\left\vert \langle \lambda _{1}\lambda _{2}\rangle _{\rho
}^{(a_{1},b_{1})}+\langle \lambda _{1}\lambda _{2}\rangle _{\rho
}^{(a_{1},b_{2})}+\langle \lambda _{1}\lambda _{2}\rangle _{\rho
}^{(a_{2},b_{1})}-\langle \lambda _{1}\lambda _{2}\rangle _{\rho
}^{(a_{2},b_{2})}\right\vert \leq 2,  \label{9}
\end{equation}
under any generalized Alice/Bob joint measurements with outcomes $\left\vert
\lambda _{1}\right\vert ,$ $\left\vert \lambda _{2}\right\vert \leq 1$ of an
arbitrary spectral type.

\section{Bell class}

Let a DSO state $\rho $ on $\mathcal{H}\otimes \mathcal{H}$ have a density
source-operator $R$ with the special dilation property 
\begin{equation}
\mathrm{tr}_{\mathcal{H}}^{(1)}[R]=\mathrm{tr}_{\mathcal{H}}^{(2)}[R]=%
\mathrm{tr}_{\mathcal{H}}^{(3)}[R]=\rho .  \label{10}
\end{equation}%
This is, in particular, the case where a state $\rho $ is reduced from a
symmetric density operator on $\mathcal{H}\otimes \mathcal{H}\otimes 
\mathcal{H}$.

\begin{theorem}
If a DSO state $\rho $ on $\mathcal{H}\otimes \mathcal{H}$ \ has a density
source-operator with the property (\ref{10}) then this DSO state $\rho $
satisfies the perfect correlation form of the original Bell inequality [1]: 
\begin{eqnarray}
\left| \text{ }\mathrm{tr}[\rho (W_{1}\otimes W_{2})]-\mathrm{tr}[\rho
(W_{1}\otimes \widetilde{W}_{2})]\text{ }\right| \leq 1-\mathrm{tr}[\rho
(W_{2}\otimes \widetilde{W}_{2})],  \label{11} \\
\left| \text{ }\mathrm{tr}[\rho (W_{1}\otimes W_{2})]-\mathrm{tr}[\rho (%
\widetilde{W}_{1}\otimes W_{2})]\text{ }\right| \leq 1-\mathrm{tr}[\rho
(W_{1}\otimes \widetilde{W}_{1})],  \notag
\end{eqnarray}
for any bounded quantum observables $W_{1},$ $\widetilde{W}_{1},$ $W_{2},$ $%
\widetilde{W}_{2}$ on $\mathcal{H}$ with operator norms $\left\| \cdot
\right\| $ $\leq 1.$
\end{theorem}

\begin{proof}
In the inequalities (\ref{4}), (\ref{5}), specified for a Bell class DSO
state $\rho ,$ let us take a DSO $R$ with the property (\ref{10}). Then, due
to (\ref{10}), $\sigma _{R}^{(1)}=\sigma _{R}^{(3)}=\rho ,$ and, therefore,
the inequalities (\ref{4}), (\ref{5}) reduce to (\ref{11}).
\end{proof}

In view of theorem 2, we refer to a DSO state having a density
source-operator with the special dilation property (\ref{10}) as \textit{a
DSO state of the Bell class. }

The set of all Bell class DSO states on $\mathcal{H}\otimes \mathcal{H}$ is
convex and includes both separable and nonseparable states. It is easy to
verify that a separable state of the special form\footnote{%
Proved by us in [5] (sec. 3.B.1) to satisfy (\ref{11}) for any three bounded
quantum observables.}: $\sum_{m}\xi _{m}\rho _{m}\otimes \rho _{m},$ where $%
\xi _{m}>0,$ $\sum_{m}\xi _{m}=1,$ represents a Bell class DSO state. In
section 3.1, we present examples of nonseparable Bell class DSO states.

Note that, satisfying the perfect correlation form of the original Bell
inequality for any three bounded quantum observables on $\mathcal{H}$, 
\textit{a Bell} \textit{class DSO state (separable or nonseparable) does not
necessarily exhibit the perfect correlation of outcomes if one and the same
quantum observable is measured on both }"\textit{sides}"\textit{. }In case
of a dichotomic quantum observable $W_{2}$, with eigenvalues $\pm 1,$ the
latter means that a Bell class DSO\ state $\rho $ satisfies the first
inequality in (\ref{11}) even if, for this state, the correlation function $%
\mathrm{tr}[\rho (W_{2}\otimes W_{2})]\neq 1$.

From theorem 2 and the representation (\ref{8}) it follows that, under\
generalized Alice/Bob joint quantum measurements, a Bell class DSO state $%
\rho $ satisfies the relation 
\begin{eqnarray}
\left\vert \langle \lambda _{1}\lambda _{2}\rangle _{\rho
}^{(a,b_{1})}-\langle \lambda _{1}\lambda _{2}\rangle _{\rho
}^{(a,b_{2})}\right\vert &=&\left\vert \mathrm{tr}[\rho (W_{1}^{(a)}\otimes
W_{2}^{(b_{1})})]-\mathrm{tr}[\rho (W_{1}^{(a)}\otimes
W_{2}^{(b_{2})})]\right\vert  \label{12} \\
&\leq &1-\mathrm{tr}[\rho (W_{2}^{(b_{1})}\otimes W_{2}^{(b_{2})})].  \notag
\end{eqnarray}%
This relation implies that, under generalized Alice/Bob joint measurements,
a Bell class DSO state $\rho $ satisfies the perfect correlation form of the
original Bell inequality: 
\begin{equation}
\left\vert \langle \lambda _{1}\lambda _{2}\rangle _{\rho
}^{(a,b_{1})}-\langle \lambda _{1}\lambda _{2}\rangle _{\rho
}^{(a,b_{2})}\right\vert \leq 1-\langle \lambda _{1}\lambda _{2}\rangle
_{\rho }^{(b_{1},b_{2})},  \label{13}
\end{equation}%
whenever $W_{2}^{(b_{1})}=W_{1}^{(b_{1})}$, that is, if POV measures of
Alice and Bob satisfy the condition $\int_{\Lambda }\lambda
_{1}M_{1}^{(b_{1})}(d\lambda _{1})=\int_{\Lambda }\lambda
_{2}M_{2}^{(b_{1})}(d\lambda _{2}).$

\textit{The latter operator condition on POV measures does not imply the
perfect correlation of outcomes and is always satisfied under Alice and Bob
projective measurements of one and the same quantum observable on both }"%
\textit{sides}"\textit{.}

\subsection{Examples}

Consider on $\mathbb{C}^{d}\otimes \mathbb{C}^{d}$, $d\geq 2,$ Werner states
[3]: $\rho ^{(d,\Phi )}=\frac{d-\Phi }{d^{3}-d}I_{\mathbb{C}^{d}\otimes 
\mathbb{C}^{d}}+\frac{d\Phi -1}{d^{3}-d}V_{d},$ $\forall \Phi \in \lbrack
-1,1],$ widely used in quantum information processing. Represented
otherwise, Werner states have the form 
\begin{equation}
\rho ^{(d,\Phi )}=\frac{1+\Phi }{2}\frac{P_{d}^{(+)}}{r_{d}^{(+)}}+\frac{%
1-\Phi }{2}\frac{P_{d}^{(-)}}{r_{d}^{(-)}}.  \label{14}
\end{equation}%
Here: (i) $V_{d}(\psi _{1}\otimes \psi _{2}):=\psi _{2}\otimes \psi _{1}$ is
the permutation (flip) operator on $\mathbb{C}^{d}\otimes \mathbb{C}^{d}$;
(ii) $P_{d}^{(\pm )}=\frac{1}{2}(I_{\mathbb{C}^{d}\otimes \mathbb{C}^{d}}\pm
V_{d})$ are the orthogonal projections onto the symmetric (plus sign) and
antisymmetric (minus sign) subspaces of $\mathbb{C}^{d}\otimes \mathbb{C}%
^{d} $ with dimensions $r_{d}^{(\pm )}=\mathrm{tr}[P_{d}^{(\pm )}]=\frac{%
d(d\pm 1)}{2}.$ For any $d\geq 2,$ a Werner state $\rho ^{(d,\Phi )}$ is
separable if $\Phi \in \lbrack 0,1]$ and nonseparable otherwise.

\begin{theorem}
\textrm{(a)} For a dimension $d\geq 3,$ every Werner state $\rho ^{(d,\Phi
)},$ separable or nonseparable, is a Bell class DSO state. \textrm{(b)} A
two-qubit Werner state $\rho ^{(2,\Phi )}$ is a Bell class DSO state
whenever $\Phi \in \lbrack 0,1]$, that is, if $\rho ^{(2,\Phi )}$ is
separable.
\end{theorem}

\begin{proof}
Introduce on $\mathbb{C}^{d}\otimes \mathbb{C}^{d}\otimes \mathbb{C}^{d}$, $%
\forall d\geq 2,$ the orthogonal projections\footnote{$Q_{d}^{(+)}$ is the
projection on the symmetric subspace of $\mathbb{C}^{d}\otimes \mathbb{C}%
^{d}\otimes \mathbb{C}^{d}.$}: 
\begin{eqnarray}
Q_{d}^{(\pm )}(\psi _{1}\otimes \psi _{2}\otimes \psi _{3}) &:&=\frac{1}{6}%
\{\psi _{1}\otimes \psi _{2}\otimes \psi _{3}\pm \psi _{2}\otimes \psi
_{1}\otimes \psi _{3}\pm \psi _{1}\otimes \psi _{3}\otimes \psi _{2}
\label{15} \\
&&\pm \psi _{3}\otimes \psi _{2}\otimes \psi _{1}+\psi _{2}\otimes \psi
_{3}\otimes \psi _{1}+\psi _{3}\otimes \psi _{1}\otimes \psi _{2}\},\text{ }
\notag
\end{eqnarray}%
$\forall \psi _{1},\psi _{2},\psi _{3}\in \mathbb{C}^{d}.$ These projections
have the form 
\begin{eqnarray}
6Q_{d}^{(\pm )} &=&I_{\mathbb{C}^{d}\otimes \mathbb{C}^{d}\otimes \mathbb{C}%
^{d}}\pm V_{d}\otimes I_{\mathbb{C}^{d}}\pm I_{\mathbb{C}^{d}}\otimes
V_{d}\pm (I_{\mathbb{C}^{d}}\otimes V_{d})(V_{d}\otimes I_{\mathbb{C}%
^{d}})(I_{\mathbb{C}^{d}}\otimes V_{d})  \label{16} \\
&&+(I_{\mathbb{C}^{d}}\otimes V_{d})(V_{d}\otimes I_{\mathbb{C}%
^{d}})+(V_{d}\otimes I_{\mathbb{C}^{d}})(I_{\mathbb{C}^{d}}\otimes V_{d}). 
\notag
\end{eqnarray}%
Taking into account that\footnote{%
Here, we use: $V_{d}^{2}=I_{\mathbb{C}^{d}\otimes \mathbb{C}^{d}},$ $\mathrm{%
tr}_{\mathbb{C}^{d}}^{(j)}[V_{d}]=I_{\mathbb{C}^{d}},$ $\mathrm{tr}_{\mathbb{%
C}^{d}}^{(j)}[P_{d}^{(\pm )}]=\frac{d\pm 1}{2}I_{\mathbb{C}^{d}},$ $\forall $
$j=1,2.$} 
\begin{eqnarray}
I_{\mathbb{C}^{d}\otimes \mathbb{C}^{d}} &=&\mathrm{tr}_{\mathbb{C}%
^{d}}^{(j)}[V_{d}\otimes I_{\mathbb{C}^{d}}]=\mathrm{tr}_{\mathbb{C}%
^{d}}^{(k)}[I_{\mathbb{C}^{d}}\otimes V_{d}]=\mathrm{tr}_{\mathbb{C}%
^{d}}^{(m)}[(I_{\mathbb{C}^{d}}\otimes V_{d})(V_{d}\otimes I_{\mathbb{C}%
^{d}})(I_{\mathbb{C}^{d}}\otimes V_{d})],\text{ }  \label{17} \\
\forall \text{\ }j &=&1,2,\text{\ \ }\forall k=2,3,\text{ \ }\forall m=1,3; 
\notag \\
V_{d} &=&\frac{1}{d}\mathrm{tr}_{\mathbb{C}^{d}}^{(3)}[V_{d}\otimes I_{%
\mathbb{C}^{d}}]=\frac{1}{d}\mathrm{tr}_{\mathbb{C}^{d}}^{(1)}[I_{\mathbb{C}%
^{d}}\otimes V_{d}]=\frac{1}{d}\mathrm{tr}_{\mathbb{C}^{d}}^{(2)}[(I_{%
\mathbb{C}^{d}}\otimes V_{d})(V_{d}\otimes I_{\mathbb{C}^{d}})(I_{\mathbb{C}%
^{d}}\otimes V_{d})]  \notag \\
&=&\mathrm{tr}_{\mathbb{C}^{d}}^{(n)}[(I_{\mathbb{C}^{d}}\otimes
V_{d})(V_{d}\otimes I_{\mathbb{C}^{d}})]=\mathrm{tr}_{\mathbb{C}%
^{d}}^{(n)}[(V_{d}\otimes I_{\mathbb{C}^{d}})(I_{\mathbb{C}^{d}}\otimes
V_{d})],\text{\ \ \ }\forall n=1,2,3,  \notag
\end{eqnarray}%
we derive 
\begin{equation}
\mathrm{tr}_{\mathbb{C}^{d}}^{(j)}[Q_{d}^{(\pm )}]=\frac{d\pm 2}{6}(I_{%
\mathbb{C}^{d}\otimes \mathbb{C}^{d}}\pm V_{d})=\frac{d\pm 2}{3}P_{d}^{(\pm
)},\ \text{\ \ }\forall j=1,2,3,  \label{18}
\end{equation}%
and $\mathrm{tr}[Q_{d}^{(\pm )}]=\frac{d(d\pm 1)(d\pm 2)}{6}.$

Consider on $\mathbb{C}^{d}\otimes \mathbb{C}^{d}\otimes \mathbb{C}^{d}$, $%
\forall d\geq 3,$ the density operator 
\begin{equation}
R^{(d,\Phi )}=\frac{1+\Phi }{2}\frac{6Q_{d}^{(+)}}{d(d+1)(d+2)}+\frac{1-\Phi 
}{2}\frac{6Q_{d}^{(-)}}{d(d-1)(d-2)}.  \label{19}
\end{equation}
Due to (\ref{14}), (\ref{18}), $\mathrm{tr}_{\mathbb{C}^{d}}^{(j)}[R^{(d,%
\Phi )}]=\rho ^{(d,\Phi )},$ for any $j=1,2,3.$ Therefore, for any state $%
\rho ^{(d,\Phi )},$ with $d\geq 3$ and $\Phi \in \lbrack -1,1],$ the
operator $R^{(d,\Phi )}$ is a density source-operator (DSO) with the
property (\ref{10}). This proves statement (\textrm{a}).

For a state $\rho ^{(2,\Phi )}=\frac{1-\Phi }{2}I_{\mathbb{C}^{2}\otimes 
\mathbb{C}^{2}}+\frac{2\Phi -1}{3}P_{2}^{(+)}$, consider the operator 
\begin{equation}
\widetilde{R}^{(2,\Phi )}=\frac{1-\Phi }{4}I_{\mathbb{C}^{2}\otimes \mathbb{C%
}^{2}\otimes \mathbb{C}^{2}}+\frac{2\Phi -1}{4}Q_{2}^{(+)}.  \label{20}
\end{equation}
This operator satisfies the relation $\mathrm{tr}_{\mathbb{C}^{d}}^{(j)}[%
\widetilde{R}^{(2,\Phi )}]=\rho ^{(2,\Phi )},$ for any $j=1,2,3,$ and is
positive whenever $\Phi \in \lbrack 0,1].$ Hence,\ for any state $\rho
^{(2,\Phi )},$ with $\Phi \in \lbrack 0,1]$, the operator $\widetilde{R}%
^{(2,\Phi )}$ represents a DSO with the property (\ref{10}). The latter
proves statement (\textrm{b}).
\end{proof}

In view of theorems 2, 3, for any dimension $d\geq 3,$ every Werner state,
separable or nonseparable, satisfies the perfect correlation form of the
original Bell inequality for any three quantum observables on\textit{\ }$%
\mathbb{C}^{d}$. For $d=2,$ the original Bell inequality holds for any
separable Werner state.

\end{document}